\documentclass[12pt]{article}
\usepackage{a4wide}
\usepackage[normalem]{ulem}
\usepackage[dvipdfmx]{graphicx}
\usepackage[top=3.5cm,bottom=3.5cm,left=2cm,right=2cm]{geometry}
\usepackage{amsmath,amssymb}
\usepackage{cancel}
\usepackage{mathtools}
\usepackage{cases}
\usepackage{braket}

\usepackage{color}

\usepackage{slashed}
\usepackage[bookmarks=true,bookmarksnumbered=true,setpagesize=false]{hyperref}

\usepackage{comment}

\newcommand{\tx}{\text}

\newcommand{\nn}{\nonumber\\}

\newcommand{\be}{\begin{equation}}
\newcommand{\e}{\end{equation}}
\newcommand{\aln}[1]{\begin{align}#1\end{align}}

\newcommand{\p}{\partial}

\newcommand{\ov}{\over}

\begin{document}
\title{
\vspace{-2cm}
\vbox{
\baselineskip 14pt
\hfill \hbox{\normalsize OU-HET 1176}\\
\mbox{}
} 
\bf \Large  
Quantum phase transition and absence of quadratic divergence
 in generalized quantum field theories  
}

\author{
Hikaru~Kawai,\thanks{E-mail: \tt hikarukawai@phys.ntu.edu.tw}{}\ {}
Kiyoharu Kawana,\thanks{E-mail: \tt kkiyoharu@kias.re.kr}{}\ {}
Kin-ya Oda,\thanks{E-mail: \tt odakin@lab.twcu.ac.jp}{}\ {} and
Kei Yagyu\thanks{E-mail: \tt yagyu@het.phys.sci.osaka-u.ac.jp}
\bigskip\\
\normalsize
 \it
 $^{*}$Department of Physics and Center for Theoretical Physics,
 \\
 \normalsize
\it 
  National Taiwan University, Taipei 106, Taiwan
  \\
 \normalsize 
 \it 
 $^{*}$Physics Division, National Center for Theoretical Sciences, Taipei 10617, Taiwan 
 \\ 
\normalsize 
\it  
$^{\dagger}$School of Physics, KIAS, Seoul 02455, Korea
\\
\normalsize
\it  $^{\ddag}$Department of Mathematics, Tokyo Woman’s Christian University, Tokyo 167-8585, Japan\\
\normalsize
\it $^{\S}$Department of Physics, Osaka University, Toyonaka, Osaka 560-0043, Japan
}

\date{\today}

\maketitle

\vspace{-1cm}
\begin{abstract}\noindent
In ordinary thermodynamics, around first-order phase transitions, the intensive parameters such as temperature and pressure are automatically fixed to the phase transition point when one controls the extensive parameters such as total volume and total energy. 
From the microscopic point of view, the extensive parameters are more fundamental than the intensive parameters.
%
Analogously, in conventional quantum field theory (QFT), coupling constants (including masses) in the path integral correspond to intensive parameters in the partition function of the canonical formulation.
Therefore, it is natural to expect that in a more fundamental formulation of QFT, coupling constants are 
dynamically fixed \emph{a posteriori}, just as the intensive parameter in the micro-canonical formulation. 
Here, we demonstrate that the automatic tuning of the coupling constants is realized at a quantum-phase-transition point at zero temperature, even when the transition is of higher order, due to the Lorentzian nature of the path integral. 
This naturally provides a basic foundation for the multi-critical point principle. 
As a concrete toy model for solving the Higgs hierarchy problem, we study how the mass parameter is fixed in the $\phi^4$ theory at the one-loop level in the micro-canonical or further generalized formulation of QFT. 
We find that there are two critical points for the renormalized mass: zero and of the order of ultraviolet-cutoff. 
In the former, the Higgs mass is automatically tuned to be zero and thus its fine-tuning problem is solved. 
We also show that the quadratic divergence is absent in a more realistic two-scalar model that realizes the dimensional transmutation. 
Additionally, we explore the possibility of fixing quartic coupling in $\phi^4$~theory 
and find that it can be fixed to a finite value. 
\end{abstract}
\newpage
\tableofcontents
\newpage

\section{Introduction and Outline}

By the discovery of the Higgs boson at LHC, it has been confirmed that the electroweak symmetry breaking is triggered by the Higgs mechanism. 
However, the question of why there is a huge hierarchy between the electroweak scale $10^{2}$ GeV and the Planck scale $10^{18}$ GeV is not unveiled yet. 
In the Standard Model (SM), the electroweak scale is merely obtained by fine-tuning the mass parameter of the Higgs potential, which makes the Higgs boson mass sensitive to ultraviolet (UV) scales such as the grand unification or Planck scale. 
Although supersymmetry has been discussed as one of the most promising new physics scenarios beyond the SM because of the stabilization of the Higgs boson mass to be the electroweak scale, it does not explain the smallness of the electroweak scale, nor is found near the expected TeV scale. 
In order to confront this fine-tuning problem, more radical and fundamental approaches beyond ordinary quantum field theory (QFT) appears to be required~\cite{Froggatt:1995rt,Froggatt:2001pa,Nielsen:2012pu,Kawai:2011qb,Kawai:2013wwa,Hamada:2014ofa,Hamada:2014xra,Hamada:2015dja,Giudice:2021viw,Jung:2021cps,Sugawara:1982fx,Strominger:1982xu,Nambu:1987hp,Coleman:1988tj}. 

Nature might already give us some important clues to tackle this problem. 
Let us recall a basic notion of statistical mechanics: there are several different formulations depending on which parameters to use as control   parameters, and they are equivalent in the thermodynamic limit $V\rightarrow \infty$. 
In particular, the most fundamental formulation is obtained with the micro-canonical ensemble where all the extensive parameters, e.g., energy $E$, volume $V$, number of particles $N$ are chosen as control parameters, while intensive parameters, e.g., temperature $T$, pressure $p$, and chemical potential $\mu$, are determined as functions of these extensive parameters in the thermodynamic limit. 
Here, the important point is that this correspondence is not injective when a system undergoes phase transitions. 
For example, in a first-order phase transition, the temperature stays at the critical temperature $T_{\rm cri}^{}(E)$ until the system releases or absorbs all the latent heat, which means that the critical point spans the finite region of $E$ in the micro-canonical formulation.\footnote{Although the terminology ``critical point'' is often used to express the end point of a phase equilibrium curve such as a vapor-liquid critical point, we do not use the term in this sense here, but use it in the sense that a phase transition of any order occurs.}
%
%
More generally, with a finite probability, intensive parameters are fixed at the point where the extensive quantities become discontinuous. 
%

In this paper, we explore a correspondence in QFT similar to that observed in statistical mechanics. Our objective is to address the fine-tuning problem by determining parameters in the canonical partition function of QFT from the micro-canonical partition function, thereby eliminating the need for tuning. In essence, the fine-tuning problem is resolved when a finite region in the parameter space of micro-canonical (or further generalized) QFT corresponds to a specific point in the canonical-QFT parameter space that appears fine-tuned. We also acknowledge various attempts to construct micro-canonical QFT, as discussed in previous studies~\cite{Froggatt:1995rt,Froggatt:2001pa,Nielsen:2012pu,Sugawara:1982fx,Strominger:1982xu,Nambu:1987hp,Coleman:1988tj} and the references therein.   
%
%
%

The concept of generalized QFT is both fascinating and intriguing; however, the actual fine-tuning of couplings remains unclear, necessitating further research.      
As a first step, we investigate the free scalar theory to explore how the bare mass-squared parameter $m_\tx{B}^2$ is fixed in the generalized QFT.    
We calculate the generalized partition function and find two critical points in the large volume limit: $m_\tx{B}^2=0$ and  ${\cal O}(\Lambda^2)$, where $\Lambda$ denotes the cut-off scale.
In particular, while the latter corresponds to a saddle point of the vacuum energy and depends on the regularization schemes in general, the former corresponds to its discontinuity and does not depend on the regularization schemes.
In this regard, one might find $m_\tx{B}^2=0$ physically more preferable, meaning that massless theory is naturally realized in the micro-canonical (or further generalized) picture.
This can be also interpreted as a theoretical explanation of the origin of the so-called \emph{classical scale invariance} or \emph{classical conformality} in the literature~\cite{Bardeen:1995kv,Meissner:2006zh,Meissner:2008gj,Foot:2007iy,Iso:2009ss,Iso:2009nw,Hur:2011sv,Iso:2012jn,Englert:2013gz,Hashimoto:2013hta,Holthausen:2013ota,Hashimoto:2014ela,Kubo:2014ova,Endo:2015ifa,Kubo:2015cna,Jung:2019dog}, which is also the basic assumption behind the Coleman-Weinberg mechanism~\cite{Coleman:1973jx}. 

We proceed to examine the $\phi^4$ theory at the one-loop level and generalize it to a large-$N$ model. 
In both cases, the mass term receives UV divergent corrections $\delta m_{\rm UV}^2\sim \Lambda^2$, prompting us to investigate at which value the renormalized mass $m^2:=m_\tx{B}^2+\delta m_{\rm UV}^2$ is fixed in the generalized QFT. 
A key distinction from the above free theory lies in the vacuum transition at $m^2=0$; 
For $m^2>0$, we observe a trivial vacuum expectation value (VEV) $\langle \phi\rangle=0$, while it becomes nonzero for $m^2<0$. 
This behavior 
corresponds to the discontinuity of the (second) derivative of the vacuum energy at $m^2=0$, which remains a critical point in both cases. 
%
Although our analysis is limited to the one-loop level or the large $N$ limit, we anticipate that the conclusions will remain unaffected by higher order corrections 
as long as the theory is in the perturbative region. 
%

In our next non-trivial example, we examine a two-real-scalar model at the one-loop level.
This model has been extensively studied in a phenomenological context~\cite{Haruna:2019zeu,Hamada:2020wjh,Kannike:2020qtw,Kawai:2021lam,Hamada:2021jls,Hamada:2022soj} because it can realize the Coleman-Weinberg mechanism~\cite{Coleman:1973jx} under the assumption of classical conformality, i.e., $m^2=0$.
However, in the generalized QFT, this is not an assumption but a point of discussion for determining where and how renormalized masses are fixed.
To streamline our analysis, we assume $\mathbb{Z}_2^{}\times \mathbb{Z}_2^{}$ symmetry and focus on the coupling space where one real scalar $\phi$ can develop a nonzero VEV $\langle \phi\rangle\neq 0$ due to the radiative potential.  
We discover that the classical conformal point $m_\phi^2=m_S^2=0$ is not a critical point in the generalized partition function. Instead, we identify another critical point with $m_\phi^{2}\neq 0$ and $m_S^2=0$, which corresponds to a quantum first-order phase transition point. 
From a phenomenological perspective, this critical point may serve as an alternative possibility for a dimensional transmutation mechanism~\cite{Hamada:2021jls,Hamada:2022soj} compared to the conventional classical conformal point; see also Ref.~\cite{Kawai:2021lam} for other possible critical points without $\mathbb{Z}_2^{}$ symmetry.  

The organization of this paper is as follows:
In Sec.~\ref{sec:2}, we introduce the generalized QFT and explain how the fine-tuning of coupling constants can be automatically achieved within this framework.
We also review the standard discussion on the equivalence between different ensembles in statistical mechanics to clarify the underlying concept. 
In Sec.~\ref{sec:fixing mass in free scalar theory}, we investigate the free scalar theory in the context of generalized QFT, focusing on how the bare mass-squared parameter is fixed in the large volume limit. 
In Sec.~\ref{sec:fixing mass in interacting scalar theory}, we delve into a more non-trivial example by considering the $\phi^4$ theory, discussing how the renormalized mass-squared is fixed at the one-loop level, as well as in the large-$N$ model.  
In Sec.~\ref{quartic coupling tuning section}, we explore the possibility of automatically tuning the quartic coupling constant. 
In Sec.~\ref{sec:5}, we analyze the two-real-scalar model within a simple parameter space where only one scalar can develop a VEV.  
Finally, in Sec.~\ref{sec:6}, we summarize our result.

Throughout the paper, we work in natural units $\hbar=c=1$ and employ the metric convention $\eta_{\mu\nu}^{}={\rm diag}(-,+,\cdots,+)$.

\section{Generalized Quantum Field Theory}\label{sec:2}
In this section, we briefly review the basics of statistical mechanics and   introduce a generalized partition function in the micro-canonical formulation of QFT, and illustrate how the tuning of the coupling constants becomes possible. 
%
See e.g.\ Ref.~\cite{Nielsen:2012pu} and Appendix~D in Ref.~\cite{Hamada:2015ria} for other reviews.
\subsection{Statistical mechanics}
To clarify the idea, let us briefly recall statistical mechanics. 
When control parameters are temperature $T$ and volume $V$, 
the system is  described by the canonical partition function
\aln{
Z_{V}(T)=e^{-F_{V}^{}(T)/T}=\sum_n e^{-E_n^{}/T},
}
where $E_n^{}$ denotes the energy eigenvalue and $F_{V}^{}(T)=-T\log Z_{V}^{}(T)$ is the Helmhortz free energy.  

On the other hand, we can also consider the micro-canonical formulation where $E$ is a control parameter instead of $T$. 
The corresponding partition function, the number of states, is given by 
\aln{
\Omega_{V}(E)=\Delta E \sum_n \delta(E_n^{}-E) = e^{S_{V}^{}(E)}, \label{eq:omega}
}
where $S_{V}^{}(E)$ denotes the entropy and $\Delta E$ denotes a sufficiently small energy interval whose effects on observables vanish in the thermodynamic limit $V\rightarrow \infty$. 
In the following, we omit the subscript $V$ for simplicity. 

The equivalence between the canonical and micro-canonical formulations can be shown as
\aln{e^{-F(T)/T}
&=\sum_n e^{-E_n/T}
=\int dE\sum_n\delta (E-E_n)e^{-E/T}
=\frac{1}{\Delta E}\int dE\,e^{S(E)-E/T} \notag
\\
&=\frac{V}{\Delta E}\int d\varepsilon\,e^{V\left(s(\varepsilon)-\varepsilon/T\right)}, \label{eq:equiv}
}
where $s=S/V$ ($\varepsilon=E/V$) represents the entropy (energy) density. 
The integration is dominated by the saddle point in the thermodynamic limit $V\rightarrow \infty$ as
\aln{
Z_V^{}(T)=e^{-F(T)/T}&\approx  \frac{T(2\pi C)^{1/2}}{\Delta E}e^{V(s(\varepsilon_*)-\varepsilon_*/T)},\label{partition functions} 
\\
\therefore\
\frac{F(T)}{T}&=\frac{E_*}{T}-S(E_*)+{\cal O}(\log V),
\label{legendre transformation} 
} 
where $C=\partial E/\partial T$ is the specific heat and  $E_*=\varepsilon_*V$ is the solution of 
\aln{\frac{\partial S}{\partial E}=\frac{1}{T}.\label{E and T}
}
Eqs.~(\ref{legendre transformation}) and (\ref{E and T}) are nothing but the Legendre transformation between the free energy and the entropy in thermodynamics. 
It is also straightforward to check the equivalence of the ensemble averages of a general observable $\hat{x}$. 
The canonical ensemble average is given by
\aln{\langle \hat{x}\rangle_{\rm can}^{}=\frac{1}{Z_V^{}(T)}\sum_n x_n^{}e^{-E_n^{}/T}, 
}
where $x_n^{}=\langle E_n^{}|\hat{x}|E_n^{}\rangle$.  
This can be written as
\aln{\langle \hat{x}\rangle^{\rm can}_{T}&=\frac{1}{Z_V^{}(T)}\sum_n x_n^{}\int dE e^{-E/T}\delta(E-E_n^{})
\\
&=\frac{1}{Z_V^{}(T)\Delta E}\int dE e^{S(E)-E/T}\langle \hat{x}\rangle^{\rm mic}_E~,
\label{canonical average}
}
where
\aln{\langle \hat{x}\rangle^{\rm mic}_{E}=\frac{\sum_n x_n^{}\delta(E-E_n^{})}{\Omega_V^{}(E)}
} 
is the  micro-canonical ensemble average. 
Because the integrand in Eq.~(\ref{canonical average}) has a strong peak at $E=E_{*}^{}$ as in Eq.~(\ref{partition functions}), we obtain
\aln{\langle \hat{x}\rangle_T^{\rm can}=\langle \hat{x}\rangle^{\rm mic}_{E}
\label{xcan and xmic}
}
in the thermodynamic limit $V\rightarrow \infty$. 
In particular, during the first-order phase transition, $T$ is equal to the critical temperature $T_c^{}(E)$, but $E$ varies between the two phases. 
This means that the coexisting phases are described in the finite parameter region of the micro-canonical scheme.
%
%
In this sense, a fine-tuning to a first-order phase transition point  $T=T_c^{}(E)$ is automatically realized as long as $E$ is the finite region. 
The correspondence is summarized in Table~\ref{tab:SM}.

\begin{table}[t]  
\centering 
\begin{tabular}{c|ccccc}  
  & Parameter & Partition function & Thermodynamic function 
 \\
\hline
Canonical & $T$ & $Z(T)=\sum_n e^{-E_n^{}/T}$ & $F(T)=-T\log Z(T)$    \\
Micro-canonical & $E$ & $\Omega(E)= \Delta E\sum_n \delta (E_n^{}-E)$  & $S(E)=\log \Omega(E)$  \\
\end{tabular}
\caption{Relation between canonical and micro-canonical formulations in statistical mechanics
}
\label{tab:SM}
\end{table}

\subsection{Micro-canonical formulation of QFT}
Now let us return to QFT. 
Conventionally we start from ``canonical" partition function with the bare coupling constants $\set\lambda:=\set{\lambda_{j}^{}}_{j=0,1,2,\dots}$:
\aln{
Z_{}^{}\left(\set\lambda\right)=\int {\cal D}\phi \exp\left(i\sum_j \lambda_{j}^{} O_j^{}[\phi]\right), 
\label{canonical QFT}
}
where $O_j^{}[\phi]$ ($j=0,1,2,\dots$) denotes a spacetime integral of a local operator such as $\int d^dx{1\ov2}\left(\p_\mu\phi(x)\right)^2$, $\int d^dx{1\ov2}\left(\phi(x)\right)^2$, $\int d^dx{1\ov4!}\left(\phi(x)\right)^4$, etc. 
After the renormalization procedure, we can calculate physical observables finitely as functions of (renormalized) couplings. 
However, the problem is that there is no principle to pick up specific values of $\set\lambda$ theoretically, and this is the very origin of the fine-tuning problems.   
Here comes the analogy between QFT and statistical mechanics into play:  
What if we start from the ``micro-canonical" picture in QFT?

The number of states $\Omega(E)$ in statistical mechanics can naturally be promoted to the partition function in QFT as
\aln{
\Omega\!\left(\set{A}\right)=\int {\cal D}\phi\prod_j \delta\!\left(O_j[\phi]-A_j^{}\right),
\label{micro-canonical QFT}
}
where we write $\set{A}:=\set{A_j}_{j=0,1,2,\dots}$ and each $A_j^{}$ is an ``extensive parameter'' corresponding to $O_j^{}[\phi]$, which is proportional to the spacetime volume $V_d$.
By using the Fourier transform of the delta function, Eq.~(\ref{micro-canonical QFT}) can be written as
\aln{
\Omega\!\left(\set{A}\right)
=\int \left(\prod_{j} d\lambda_{j}\right) \int {\cal D}\phi\exp\left(i\sum_l \lambda_{l}^{} \left(O_l^{}[\phi]-A_l^{}\right)\right)
=\int \left(\prod_j d\lambda_j\right) Z\!\left(\set\lambda\right) \,e^{-i\sum_l\lambda_{l}^{}A_l^{}}.
	\label{microcanonical QFT}
}
%
Now, one can see that we have an ensemble average of various coupling constants and their weight is proportional to the canonical partition function~\eqref{canonical QFT}.

\subsection{Further generalized QFT}
Further, we can generalize~\cite{Nielsen:2012pu} the product of delta functions in Eq.~\eqref{micro-canonical QFT} to an arbitrary weight function~$W\!\left(\set{x 
}\right)$ as~\footnote{
Here, it is understood that $\set{x
}:=\set{x_j^{}
}_{j=0,1,2,\dots}$.
}
\aln{
\Omega\!\left(\set{A}\right)&=\int {\cal D}\phi\,W\!\left(\set{O[\phi]-A}\right)
\notag\\
&=\int\left(\prod_j d\lambda_{j}\right) Z\!\left(\set\lambda\right)e^{-i\sum_l\lambda_{l}^{}A_l^{}}\omega\!\left(\set\lambda\right),
\label{generalized QFT}
}
where $\omega\!\left(\set\lambda\right)$ is the Fourier transform of $W\!\left(\set{x}\right)$:
\aln{
W\!\left(\set{x
}\right)
	&=	\int\left(\prod_jd\lambda_j\,e^{i\lambda_j x_j^{}
	}\right)\omega\!\left(\set\lambda\right). 
}
In this generalization, we assume that $W\!\left(\Set{x}\right)$ can depend on extensive parameters solely through $x_j^{}={\cal O}_j^{}[\phi]-A_j^{}$
so that $\omega\!\left(\Set{\lambda}\right)$ does not depend on any extensive parameters.
%

If there exists a strong peak $\set{\lambda^{*}}$ of $Z\!\left(\set{\lambda}\right)$ in the infinite volume limit $V\rightarrow \infty$, the generalized QFT is equivalent to the ordinary canonical QFT whose (bare) coupling constants are fixed at $\{\lambda^{*}\}$. 
%
In other words, the fine-tuning of coupling constants is automatically realized in the generalized partition function.  
We will see that only the behavior of the \emph{canonical} partition function $Z\!\left(\Set{\lambda}\right)$ matters for the realization of the Higgs fine-tuning, regardless of whether it is in the micro-canonical QFT~\eqref{micro-canonical QFT} or in the generalized QFT~\eqref{generalized QFT}.
See Table~\ref{tab:QFT} for the summary of naive correspondence. 
In the following sections, we will verify this fine-tuning mechanism for the (bare) mass term $m_\tx{B}^2$ in scalar field theory. 

\begin{table}[t]  
\centering 
\begin{tabular}{c|ccccccc}  
  & Parameter & Partition function & Generating function 
 \\
\hline
Canonical QFT & $\lambda$ & $Z(\lambda)=\int {\cal D}\phi e^{i\lambda O[\phi]}$ & $F(\lambda)=-i^{-1}\log Z(\lambda)$    \\
Micro-canonical QFT & $A$ & $\Omega(A)=\int {\cal D}\phi\,\delta\!\left(O[\phi]-A\right)$  & $S(A)=\log \Omega(A)$ \\
Generalized QFT & $A$ & $\Omega(A)=\int {\cal D}\phi\,W\!\left(O[\phi]-A\right)$ & $S(A)=\log \Omega(A)$
\end{tabular}
\caption{Naive correspondence in QFT }
\label{tab:QFT}
\end{table}


\section{Fixing mass in free scalar theory}\label{sec:fixing mass in free scalar theory}

We study the free scalar theory in the generalized QFT and show how the bare mass parameter is fixed.   
%
%

\subsection{Partition function of free scalar theory} \label{sec:pf}
We consider the free scalar theory in the $d$-dimensional spacetime:
\aln{
S_0^{}=-\int d^dx\frac{1}{2}\left(\partial_\mu^{}\phi\right)^2.
	\label{S0}
}
Here, we introduce the bare mass term according to the generalized QFT, while leaving the kinetic term~\eqref{S0} as is, for simplicity.   
Then the generalized partition function is defined by
\aln{
\Omega(A)&:=\int {\cal D}\phi e^{iS_0^{}}W\!\left(A-\frac{1}{2}\int d^dx \phi^2(x)\right)
\\
&=\int_{-\infty}^{\infty} \frac{dm_\tx{B}^2}{2\pi} \int {\cal D}\phi \omega\!\left(m_\tx{B}^2\right)\exp \left\{i\left[ S_0^{} + m_\tx{B}^2\left(A-\frac{1}{2}\int d^dx \phi^2(x)\right)\right]\right\}
\\
&=\int_{-\infty}^{\infty} \frac{dm_\tx{B}^2}{2\pi}\omega\!\left(m_\tx{B}^2\right) e^{im_\tx{B}^2A}Z\!\left(m_\tx{B}^2\right), \label{eq:omega0}
}
where $\omega\!\left(m_\tx{B}^2\right)$ is the Fourier transform of $W(x)$ and 
$Z\!\left(m_\tx{B}^2\right)$ denotes the ordinary canonical partition function
\aln{Z\!\left(m_\tx{B}^2\right)= \int {\cal D}\phi \exp\left[i\left(S_0^{} - \frac{m_\tx{B}^2}{2}\int d^dx \phi^2(x)\right)\right].
}
We assume that $\omega\!\left(m_\tx{B}^2\right)$ is an ordinary smooth function. 
%
We can now perform the path-integral as
\aln{
\Omega(A) &\propto \int_{-\infty}^{\infty} \frac{dm_\tx{B}^2}{2\pi}\omega\!\left(m_\tx{B}^2\right)\exp \left[im_\tx{B}^2 A-\frac{1}{2}\text{tr}\log\left[i\left(-\Box + m_\tx{B}^2 - i\varepsilon\right)\right]\right]
\\
&=
\int_{-\infty}^{\infty} \frac{dm_\tx{B}^2}{2\pi}\omega\!\left(m_\tx{B}^2\right)\exp\left[iV_d^{}\left(a m_\tx{B}^2 - F\!\left(m_\tx{B}^2\right)\right)\right],
\label{free micro canonical}
}
where $\Box=\eta^{\mu\nu}\partial_\mu^{}\partial_\nu^{}$, $a=A/V_d^{}$, $i\varepsilon$ is the Feynman's prescription, and 
\aln{
F\!\left(m_\tx{B}^2\right)=\frac{1}{2}\int \frac{d^dp_\tx{E}^{}}{(2\pi)^d}\log\!\left(p_\tx{E}^2+m_\tx{B}^2-i\varepsilon\right) +\rm const., 
\label{f}
}
with $p_\tx{E}^\mu$ being the Euclidean momentum. 
This is apparently UV divergent and we need a regularization. 

\subsection{Mass-squared in free scalar theory}\label{Mass-squared in free scalar theory}
To discuss how the mass is tuned, we employ cut-off and dimensional regularizations.

\subsubsection*{Cut-off regularization}

\begin{figure}
\begin{center}\hfill
\includegraphics[width=8cm]{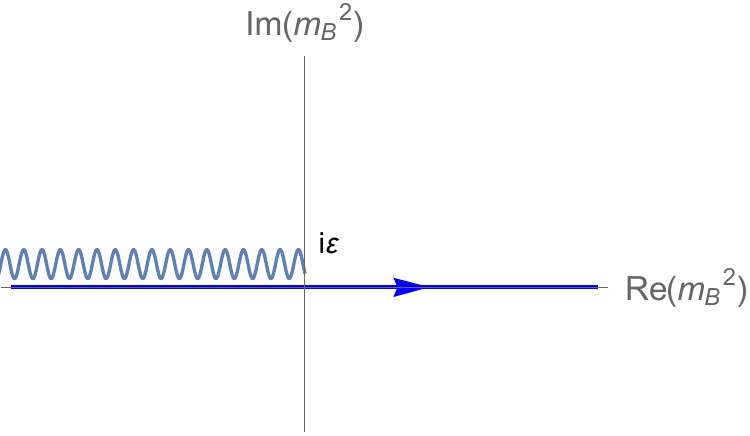}\hfill
\includegraphics[width=8cm]{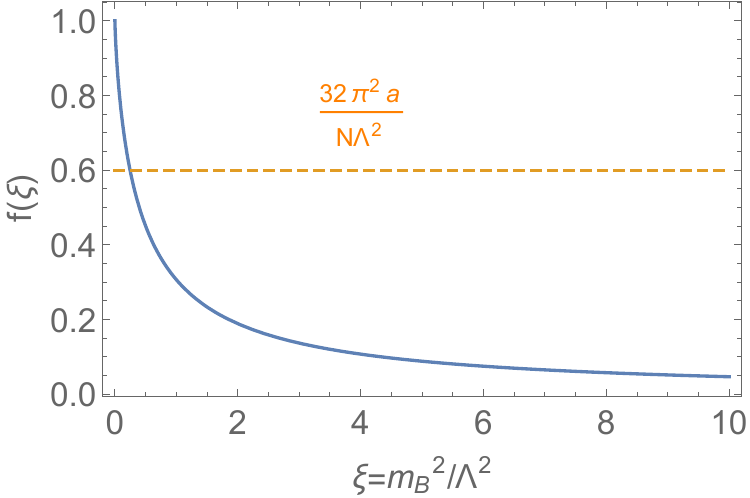}\hfill\mbox{}
\end{center}
\caption{Left: Integration line. Right: $f(\xi)$ (blue) and $32\pi^2a/(N\Lambda^2)$ (orange).  
}
\label{fig:contour}
\end{figure}

First, let us study the cut-off regularization. 
%
As a function of $m_\tx{B}^2$, the integrand in Eq.~(\ref{free micro canonical}) has a branch cut at 
\aln{
-\Lambda^2+i\varepsilon \leq  m_\tx{B}^2\leq i\varepsilon, 
\label{branch cut}
}
and the integration line $-\infty<m_\tx{B}^2<+\infty$ is located below this branch cut as shown in the left panel in Fig.~\ref{fig:contour}.     
Then, Eq.~(\ref{f}) can be evaluated as 
\aln{
F\!\left(m_\tx{B}^2\right)&=\frac{S_{d-1}^{}}{2(2\pi)^d}\int_0^\Lambda dp_\tx{E}^{}p_\tx{E}^{d-1}\log\!\left(p_\tx{E}^2+m_\tx{B}^2-i\varepsilon\right)
\\
&=\frac{S_{d-1}^{}\Lambda^{d}}{4(2\pi)^d}\int_0^1 dx x^{\frac{d}{2}-1}\log\!\left(x+\xi-i\varepsilon\right),\quad \xi=m_\tx{B}^2/\Lambda^2
\\
&=\frac{S_{d-1}^{}\Lambda^{d}}{4(2\pi)^d}\times \begin{cases}\int_0^1 dx x^{\frac{d}{2}-1}\log(x+\xi) & \text{for }\xi>0
\\
\int_0^1 dx x^{\frac{d}{2}-1}\log|x+\xi|-i\frac{2\pi}{d}(-\xi)^{\frac{d}{2}} & \text{for }0>\xi>-1
\\
\int_0^1 dx x^{\frac{d}{2}-1}\log|x+\xi|-i\frac{2\pi}{d} & \text{for }-1>\xi
\end{cases}, 
}
where $S_{d-1}^{}$ denotes the area of a $d-1$ dimensional sphere.

We see that the function $F(m_\tx{B}^2)$ contains the imaginary part for $m_\tx{B}^2<0$, which gives a large suppression in the  partition function when $V_d^{}$ is large.  
Qualitatively, the contribution from $m_\tx{B}^2<0$ is 
\aln{\int^0_{-\infty} d\xi e^{-V_d^{}\Lambda^d(-\xi)^{d/2}}={\cal O}\left(\frac{1}{(V_d^{}\Lambda^d)^{\frac{2}{d}}}\right). 
}
 
%
On the other hand, there is no such suppression for $m_\tx{B}^2>0$, and the integrand is a rapidly oscillating function of $m_\tx{B}^2$. 
In this case, the saddle point can exist at the point determined by
\aln{
a&=\frac{dF}{dm_\tx{B}^2}=\frac{S_{d-1}^{}\Lambda^{d-2}}{4(2\pi)^{d}}\int_0^1 dx\frac{x^{\frac{d}{2}-1}}{x+\xi},
\label{saddle point of mass}
}
When $d=4$, this equation becomes 
\aln{
a=\frac{S_{3}^{}\Lambda^2}{4(2\pi)^{4}}\left(1-\xi\log\frac{1+\xi}{\xi}\right)=:\frac{\Lambda^2}{32\pi^{2}}f(\xi),
\label{free sadlle point}
}
where $f(\xi)$ is a monotonic function satisfying $f(0)=1$ and $f(\infty)=0$; see the right panel in Fig.~\ref{fig:contour}.  
Thus, there can be a saddle point when $a/\Lambda^2\leq \left(32\pi^2\right)^{-1}$, while the exponent, $a m_\tx{B}^2 - F\!\left(m_\tx{B}^2\right)$, becomes a monotonic function when $a/\Lambda^2> \left(32\pi^2\right)^{-1}$. 
%
%

\subsubsection*{Dimensional regularization}
Let us also calculate the partition function in the dimensional regularization. 
We define $\epsilon=4-d$. 
In this case, the free energy is calculated as
\aln{
F\!\left(m_\tx{B}^2\right) &=-\frac{\left(m_\tx{B}^2\right)^2}{2(4\pi)^2}\left(\frac{1}{\epsilon}-\frac{\gamma}{2}+\frac{3}{4}+\frac{1}{2}\log 4\pi-\frac{1}{2}\log \left(\frac{m_\tx{B}^2-i{\varepsilon}}{\mu^2}\right)\right)
\notag\\
&=:-\frac{\left(m_\tx{B}^2\right)^2}{2(4\pi)^2}\left(c_{\overline{\rm MS}}^{}-\frac{1}{2}\log \left(\frac{m_\tx{B}^2-i{\varepsilon}}{\mu^2e^{3/2}}\right)\right),
\label{dimensional regularization 1}
}
where $\mu$ is the renormalization scale and $c_{\overline{\rm MS}}^{}$ contains both the finite and divergent terms. 

For $m_\tx{B}^2\leq 0$, we again have the imaginary part from the logarithmic term, and the partition function is highly suppressed. 
On the other hand, for $m_\tx{B}^2>0$, the saddle point is determined by
\aln{a=\frac{dF}{dm_\tx{B}^2}=\frac{m_\tx{B}^2}{32\pi^2}\log\left(\frac{m_\tx{B}^2}{\mu^2 e^{1+2c_{\overline{\rm MS}}^{}}}\right),
} 
which always has a solution 
unlike the cut-off case. 
If we identify $\mu e^{c_{\overline{\rm MS}}^{}}$ as the cut-off scale, 
the location of the saddle points is   
\aln{
m_\tx{B}^2\sim \Lambda^2\sim \mu^2 e^{2c_{\overline{\rm MS}}^{}}~
}
in both of the regularizations. 
%
%

\subsubsection*{Mass tuning in free scalar theory}
As we have seen, the existence of a saddle point depends on the regularization scheme. 
But in any case, $m_\tx{B}^2=0$ holds over a wide range of the parameter space. 
First, if there is no saddle point, the $m_\tx{B}^2$ integration is dominated by the boundary $m_\tx{B}^2=0$ by the mathematical formula (\ref{kink delta}) for any smooth weight function $\omega(m_\tx{B}^2)$. 
On the other hand, if there is a saddle point, $am_\tx{B}^2-F(m_\tx{B}^2)$ is monotonic below the saddle point $m_\tx{B}^2\lesssim\Lambda^2$. 
Thus, when the weight function $\omega\!\left(m_\tx{B}^2\right)$ has a finite support for $m_\tx{B}^{2}\lesssim \Lambda^2$, the free energy $am_\tx{B}^2-F\!\left(m_\tx{B}^2\right)$ is monotonic, and the boundary $m_\tx{B}^2=0$ is always dominant due to the mathematical formula~(\ref{kink delta}). 
%
%
In conclusion, $m_\tx{B}^2=0$ appears to be a unique critical point that is physically reasonable in the generalized partition function~(\ref{free micro canonical}).

\subsection{Equivalence in large volume limit}
Finally, let us confirm the equivalence between generalized QFT and canonical QFT in the large volume limit. 
When $A\neq0$, the first derivative of $m_\tx{B}^2A/V_d^{}-F(m_\tx{B}^2)$ is continuous and nonzero at $m_\tx{B}^2=0$ while the second derivative is discontinuous. 
Then, we can use the mathematical formula~(\ref{general discontinuity}) and $\Omega(A)$ is evaluated as  
\aln{
\lim_{V_d^{}\rightarrow \infty}\Omega(A)&=\lim_{V_d^{}\rightarrow \infty}\int_{-\infty}^\infty \frac{dm_\tx{B}^2}{2\pi} \omega(m_\tx{B}^2)e^{im_\tx{B}^2A}Z(m_\tx{B}^2)
=
{\cal N} \lim_{V_d^{}\rightarrow \infty}\frac{Z(m_\tx{B}^2=0)}{V_d^{2}}~,
}
where ${\cal N}$ is an unimportant numerical factor.
That is,
\aln{
\lim_{V_d^{}\rightarrow \infty}\log \Omega (A)=\lim_{V_d^{}\rightarrow \infty}\log Z(m_\tx{B}^2=0)+{\cal O}(\log V_d^{}).
}
We apparently see the equivalence in the large volume limit.
%

We can also check the equivalence of correlation functions. 
We introduce a source term as 
\aln{
\Omega[A;J]&=\int {\cal D}\phi e^{iS_0^{}-i\int d^dxJ(x)\phi(x)}W\!\left(A-\frac{1}{2}\int d^dx \phi^2(x)\right)
\nn
&=\int_{-\infty}^{\infty} \frac{dm_\tx{B}^2}{2\pi}\omega\!\left(m_\tx{B}^2\right)e^{iV_d^{}\left(m_\tx{B}^2 a-F(m^2)\right)} \exp\left(\frac{i}{2}\int d^dxJ(x)(-\Box+m_\tx{B}^2-i\epsilon)^{-1}J(x)\right)
\nn
&=\int_{-\infty}^{\infty} \frac{dm_\tx{B}^2}{2\pi}\omega\!\left(m_\tx{B}^2\right)e^{iAm_\tx{B}^2}
Z\!\left(m_\tx{B}^2\right) G(J;m_\tx{B}^2)~. 
\label{G}
}
As long as $J(x)$ is a finite supported function, i.e., when it has no volume dependence, the factor~$G(J;m_\tx{B}^2)$ does not have exponentially large volume dependence, and the $m_\tx{B}^2$ integral in Eq.~(\ref{G}) is dominated by the critical point of $Z\!\left(m_\tx{B}^2\right)$. 
%
%
Namely, we have 
\aln{
&\lim_{V_d^{}\rightarrow \infty} \Omega[A;J]= {\rm const.}\times 
G(J;m_\tx{B}^2=0). 
}
%
Then, by taking the functional derivatives with respect to $J(x)$, we obtain 
\aln{
\left\langle T\left\{\phi(x_1^{})\cdots \phi(x_n^{})\right\}\right\rangle_{\rm mic}^{}=\left\langle T\left\{\phi(x_1^{})\cdots \phi(x_n^{})\right\}\right\rangle_{\rm can}^{}\bigg|_{m_\tx{B}^2=0}^{}, 
} 
which corresponds to Eq.~(\ref{xcan and xmic}) in statistical mechanics.   

\section{Fixing mass in $\phi^4$ theory}\label{sec:fixing mass in interacting scalar theory}
In this section, we analyze the $\phi^4$ theory in the scope of generalized QFT. 
%
%
In particular, we study how the mass term is fixed by studying its critical point at the one-loop level, and also in the large-$N$ extended model.
%
%
Our investigation reveals that the renormalized mass settles at zero in the one-loop analysis as well as in the large-$N$ model, due to the discontinuity of the derivative of the vacuum energy.

In this section, we assume that the quartic coupling is prefixed at a certain value, as in the ordinary QFT. The possibility of its tuning will be discussed in the next section.

\subsection{Mass-squared at one-loop level}\label{one-loop}
We first introduce a bare action without the bare mass term:
\aln{
S_\tx{B}^{}[\phi]=\int d^dx \left(-\frac{1}{2}(\partial_\mu^{}\phi)^2-\frac{\lambda_\tx{B}^{}}{4!}\phi^4-\Lambda_{\rm B}^{}\right),
	\label{phi4 action}
}
where we have also included the bare cosmological constant term for later convenience. 
For simplicity, we have imposed the $\mathbb{Z}_2^{}$ symmetry $\phi\rightarrow -\phi$, which leaves only the quadratic and quartic terms in the renormalizable potential.

Our starting point is the generalized partition function $\Omega(A)$ which can be expressed in terms of the canonical partition function $Z\!\left(m_\tx{B}^2\right)$ by the same procedure as in Sec.~\ref{sec:pf}:
\aln{
\Omega(A)&=
\int_{-\infty}^{\infty} \frac{dm_\tx{B}^2}{2\pi} \omega\!\left(m_\tx{B}^2\right)e^{im_\tx{B}^2 A}\, Z\!\left(m_\tx{B}^2\right),
\label{O(N) model}
}
where
\aln{
Z\!\left(m_\tx{B}^2\right) 
	&=	\int {\cal D}\phi e^{iS[\phi]},&
S^{}[\phi]
	&=	S_\tx{B}^{}[\phi]-\frac{m_\tx{B}^2}{2}\int d^dx\phi^2.
	\label{phi4 action}
}
The canonical partition function $Z$ can be obtained from the ordinary effective action $\Gamma[\phi]$ via $Z=e^{i\min_\phi\Gamma[\phi]}$. At the one-loop level, we obtain
\aln{
\Gamma[\phi]&=S[\phi]-\frac{1}{2i}{\rm tr}\log \left(-\Box+m_\tx{B}^2+\frac{\lambda_\tx{B}^{}}{2}\phi_{}^2-i\varepsilon\right)
\notag\\
&=S[\phi]+\frac{V_d^{}}{2}\int\frac{d^dp_\tx{E}^{}}{(2\pi)^d}
\log \left(p_\tx{E}^2+m_\tx{B}^2+\frac{\lambda_\tx{B}^{}}{2}\phi_{}^2-i\varepsilon\right)
\notag\\
&=S[\phi]+\frac{V_d^{}}{2}\frac{(M_\phi^2(\phi))^2}{(4\pi)^2}\left[ c_{\overline{\rm MS}}^{}-\frac{1}{2}\log \left(\frac{M_\phi^2(\phi)-i\varepsilon}{\mu^2e^{3/2}}\right)\right],
\label{log Z}
}
%
%
where $M_\phi^{2}(\phi_{}^{})=m_\tx{B}^2+\frac{\lambda_\tx{B}^{}}{2}\phi_{}^2$. 

Now the one-loop effective potential is given by
\aln{
V_{\rm eff}^{}(\phi)
	:=	-\frac{\Gamma[\phi]}{V_d^{}}\bigg|_{\phi={\rm const}}^{}
	&=	\Lambda_\tx{B}^{}+\frac{m_\tx{B}^2}{2}\phi_{}^2+\frac{\lambda_\tx{B}^{}}{4!}\phi^4
-\frac{(M_\phi^2(\phi^{}))^2}{32\pi^2}\left[ c_{\overline{\rm MS}}^{}-\frac{1}{2}\log \left(\frac{M_\phi^2(\phi)-i\varepsilon}{\mu^2e^{3/2}}\right)\right]
\notag\\
&=\Lambda_\tx{R}^{}+\frac{m_\tx{R}^2}{2}\phi_{}^2+\frac{\lambda_\tx{R}^{}}{4!}\phi_{}^4
+\frac{(M_\phi^2(\phi^{}))^2}{64\pi^2}\log \left(\frac{M_\phi^2(\phi)-i\varepsilon}{\mu^2e^{3/2}}\right),
\label{one loop potential for O(N)}
}
where 
\aln{
\Lambda_\tx{R}^{}
	&=	\Lambda_\tx{B}^{}-c_{\overline{\rm MS}}^{}\frac{m_\tx{B}^4}{32\pi^2},&
m_\tx{R}^2
	&=	m_\tx{B}^2-\frac{c_{\overline{\rm MS}}^{}}{16\pi^2}
\lambda_\tx{B}^{}m_\tx{B}^2,&
\lambda_\tx{R}^{}
	&=	\lambda_\tx{B}^{}-\frac{3c_{\overline{\rm MS}}^{}}{16\pi^2}
\lambda_\tx{B}^2, \label{eq:parameters_ms}
}
denote the renormalized parameters. 
We can regard these renormalized couplings as our parameters instead of the bare couplings $(m_\tx{B}^2,\lambda_\tx{B}^{})$.

As a consistency check, let us consider the free theory limit: $\lambda_\tx{B}^{}=\lambda_\tx{R}^{}=0$.
In this case, there is no difference between the bare mass and the renormalized mass, and the effective potential is exactly given by
\aln{
V_{\rm eff}^{}(\phi)\bigg|_{\lambda_\tx{B}^{}=0}^{}=\Lambda_\tx{R}^{}+\frac{m_\tx{B}^2}{2}\phi^2+\frac{m_\tx{B}^4}{64\pi^2}\log\left(\frac{m_\tx{B}^2-i{\varepsilon}}{\mu^2e^{3/2}}\right),
}
which has a trivial minimum at $\phi_{}^{}=0$ and the corresponding vacuum energy is 
\aln{
V_{\rm min}^{}=\Lambda_\tx{B}^{}-\frac{\left(m_\tx{B}^2\right)^2}{32\pi^2}\left(c_{\overline{\rm MS}}^{}-\frac{1}{2}\log\left(\frac{m_\tx{B}^2-i{\varepsilon}}{\mu^2e^{3/2}}\right)\right),
}
which is nothing but Eq.~(\ref{dimensional regularization 1}).   

Let us come back to the interacting scalar theory.  
We assume $\lambda_\tx{R}^{}>0$ to ensure the stability of the effective potential. 
As usual, we can obtain the RG improved effective potential by choosing the renormalization scale $\mu $ appropriately. 
For $m_\tx{R}^2\geq 0$, the VEV is trivial $v=0$, and we take $\mu=m_{R}^{}e^{-3/4}$, which will give a simple expression of the vacuum energy (as the first line in Eq.~(\ref{one-loop vacuum energy}) below). 
On the other hand, for $m_\tx{R}^2<0$, the field $\phi$ develops a nonzero VEV, 
%
%
%
%
%
%
%
which is determined by 
\aln{
\frac{\partial V_{\rm eff}^{}(\phi)}{\partial \phi^{}}=0&\quad \Leftrightarrow \quad m_\tx{R}^2 v+\frac{\lambda_\tx{R}^{}}{6}v^3+\frac{\lambda_\tx{R}^{}v}{32\pi^2}\left(M_\phi^2(v)\log\left(\frac{M_\phi^2(v)}{\mu^2e}\right)
\right)=0~.
\label{one loop VEV}
}
By choosing the renormalization scale at $\mu^2=M_\phi^{2}(v)e^{-1}$, 
the VEV is given by 
\aln{
v^2 = -\frac{6m_\tx{R}^2}{\lambda_\tx{R}^{}}~.
}
%
Now the vacuum energy as a function of $m_\tx{R}^2$ is given by
\aln{
V_{\rm min}^{}(m_\tx{R}^2)=\Lambda_{\text{B}}^{}-\frac{c_{\overline{\rm MS}}}{32\pi^2}(m_\tx{R}^2)^2
+\begin{cases}
0 & \text{for }m_\tx{R}^2\geq 0
\\
-\frac{3(m_\tx{R}^2)^2}{2\lambda_\tx{R}^{}}-\frac{\left(2m_\tx{R}^2\right)^2}{128\pi^2}
& \text{for }m_\tx{R}^2<0
\end{cases},
\label{one-loop vacuum energy}
}
which shows that the second derivative of $V_{\rm min}^{}(m_\tx{R}^2)$ is discontinuous at $m_\tx{R}^2=0$ as
\aln{
\frac{\partial^2 V_{\rm min}^{}}{\partial (m_\tx{R}^2)^2}\bigg|_{m_\tx{R}^2=0+}^{}
	&=	-\frac{c_{\overline{\rm MS}}^{}}{16\pi^2},&
\frac{\partial^2 V_{\rm min}^{}}{\partial (m_\tx{R}^2)^2}\bigg|_{m_\tx{R}^2=0-}^{}
	&\simeq
		-\frac{c_{\overline{\rm MS}}^{}}{16\pi^2}-\frac{3}{\lambda_\tx{R}^{}}+\frac{1}{16\pi^2}
		~.
}
%
Note that the first derivative of $m_\tx{B}^2A/V_d^{}-V_{\rm min}^{}(m_\tx{R}^2)$ is continuous and nonzero  at $m_\tx{R}^2=0$ as long as $A\neq 0$, which means that we can use the mathematical formula Eq.~(\ref{general discontinuity}).
Then, 
the partition function is evaluated as
\aln{\lim_{V_d^{}\rightarrow \infty}e^{im_B^2A+\log Z(m_\tx{R}^2)}=\lim_{V_d^{}\rightarrow \infty}e^{im_B^2A-iV_d^{} V_{\rm min}^{}(m_\tx{R}^2)}
\propto \lim_{V_d^{}\rightarrow \infty}\frac{e^{-iV_d^{} V_{\rm min}^{}(m_\tx{R}^2=0)}}{V_d^{2}}\delta(m_\tx{R}^2),
}
which leads to 
\aln{\lim_{V_d^{}\rightarrow \infty}^{}\Omega (A)&=\int_{-\infty}^\infty \frac{dm_\tx{R}^{2}}{2\pi}\left(\frac{\partial m_\tx{B} ^2}{\partial m_\tx{R}^2}\right)\omega(m_\tx{B}^2)e^{im_\tx{B}^2A+\log Z(m_\tx{R}^2)}
=\frac{{\cal N}}{V_d^2} e^{\log Z(m_\tx{R}^2=0)},
}
where ${\cal N}$ is an unimportant factor.
That is,
\aln{
\lim_{V_d^{}\rightarrow \infty}\log \Omega(A)&=
\lim_{V_d^{}\rightarrow \infty}\log Z(m_\tx{R}^2=0)+{\cal O}\!\left(\log V_d^{}\right),
}
%
We see that the equivalence between two formulations still holds in the $\phi^4$ theory at the one-loop level in the large volume limit.
In particular, the renormalized mass parameter $m_\tx{R}^2$ is fixed at zero because of the discontinuity of the second derivatives of the vacuum energy. \footnote{Precisely speaking, there is also a saddle point solution $m_\tx{R}^2\sim -\frac{A}{V_d^{}c_{\overline{\rm MS}}^{}}$. 
The value of which depends on the regularization schemes and the way the large volume limit $A/(c_{\overline{\rm MS}}^{}V_d^{})=$fixed is taken.  
%
On the other hand, the critical point $m_\tx{R}^2=0$ has no such uncertainty and is uniquely determined as in the free case.   
}
This result implies that the quadratic divergence problem is absent in the micro-canonical formulation. 
Moreover, in the simple $\phi^4$ theory, the conclusion is not affected by higher-loop  corrections as long as $\lambda_\tx{R}^{}\lesssim 1$. 
However, the situation can be different in more general theories with more than one field 
 because another mass scale can be radiatively generated by other fields.  
See Sec.~\ref{sec:5} for a concrete two-scalar model. 
%

\subsection{Mass-squared in Large-$N$ model}\label{large N}
In this section, we turn to the $O(N+1)$ symmetric scalar theory with $\phi_i$ ($i=0,1,\dots,N$), and discuss how the mass parameter is fixed in the generalized partition function in the large-$N$ limit.  
The bare action is given by 
\aln{
S_\tx{B}^{}=\int d^dx \left(-\frac{1}{2}(\partial_\mu^{}\phi_i)^2-\frac{\lambda_\tx{B}^{}}{4!}(\phi_i^2)^2\right),
}
where we have omitted the bare mass term as before and apply the Einstein summation convention for the field index $i$.  
%
%
%
%

\begin{figure}
\begin{center}
\includegraphics[width=10cm]{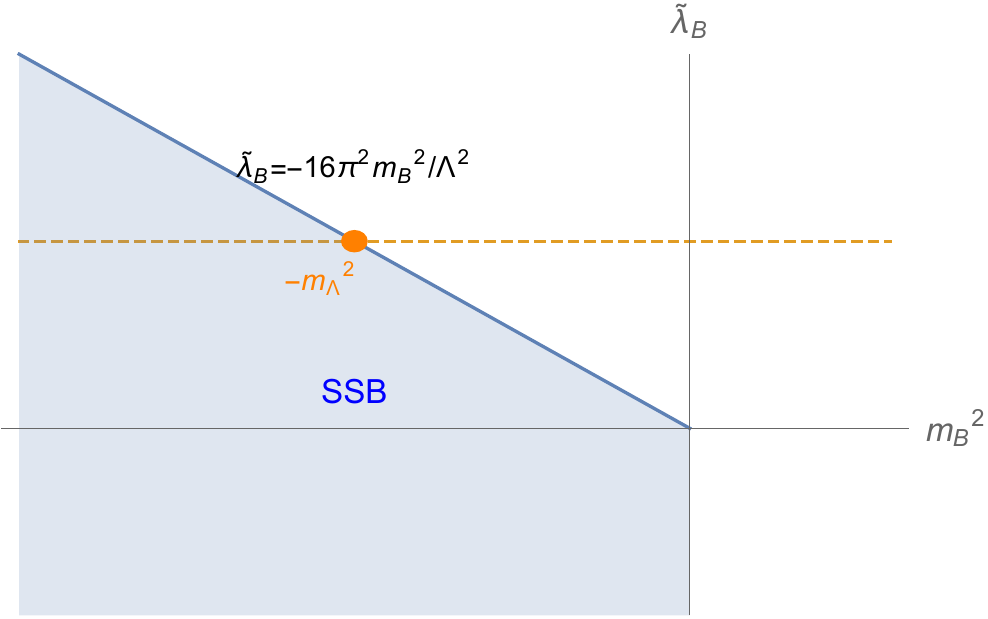}
\end{center}
\caption{Left: Parameter space in the large $N$ limit.  
}
\label{fig:large N}
\end{figure}

The generalized partition function now becomes
\aln{
\Omega(A)
=\int_{-\infty}^{\infty} \frac{dm_\tx{B}^2}{2\pi} \omega\!\left(m_\tx{B}^2\right)e^{im_\tx{B}^2 NA}\int {\cal D}\phi e^{iS^{}}
&=\int_{-\infty}^{\infty} \frac{dm_\tx{B}^2}{2\pi}  \omega\!\left(m_\tx{B}^2\right)e^{im_\tx{B}^2 NA}Z\!\left(m_\tx{B}^2\right),
\label{O(N) model}
}
where
\aln{S^{}=\int d^dx \left(-\frac{1}{2}(\partial_\mu^{}\phi_i)^2-\frac{1}{2}m_\tx{B}^2 \phi_i^2-\frac{\lambda_\tx{B}^{}}{4!}(\phi_i^2)^2
\right)~.
}
%
We can generally separate the original field $\phi_i^{}$ as
\aln{
\phi_i^{}=\begin{cases} \sqrt{N}s & i=0
\\
\pi_i^{} & i=1,2,\cdots,N
\end{cases}\quad , \label{eq:decompose}
}
where $s$ is the field that may acquire a VEV.   
The Lagrangian now becomes
\aln{
{\cal L}=N\left(-\frac{1}{2}(\partial_\mu^{}s)^2-\frac{m_\tx{B}^2}{2}s^2\right) - \frac{1}{2}(\partial_\mu^{}\pi_i^{})^2-\frac{m_\tx{B}^2}{2}\pi_i^2-\frac{\lambda_\tx{B}^{}}{4!}(Ns^2+\pi_i^2)^2.
}
We can introduce an auxiliary scalar field $c$ in the Lagrangian such that the partition function does not change after performing the path-integral over $c$: 
\aln{{\cal L}
&=N\left(-\frac{1}{2}(\partial_\mu^{}s)^2-\frac{1}{2}(m_\tx{B}^2+c)s^2\right) - \frac{1}{2}(\partial_\mu^{}\pi_i^{})^2-\frac{1}{2}(m_\tx{B}^2+c)\pi_i^2+\frac{3}{2\lambda_\tx{B}^{}}c^2. 
}
By performing the path-integral of $\pi_i^{}$, the partition function is given as 
\aln{Z(m_\tx{B}^2)=\int {\cal D}\phi e^{iS^{}}
=\int {\cal D}c\int {\cal D}s\exp\bigg[iN\int d^d x\bigg(&-\frac{1}{2}(\partial_\mu^{}s)^2-\frac{1}{2}(m_\tx{B}^2+c)s^2+\frac{1}{4\tilde{\lambda}_\tx{B}^{}}c^2
\nn
&-\frac{1}{2}\int\frac{d^dp_\tx{E}^{}}{(2\pi)^d}\log (p_\tx{E}^2+m_\tx{B}^2+c-i\epsilon)\bigg) \bigg].
\label{effective theory in large N}
}
where $\tilde{\lambda}_\tx{B}^{}:=\lambda_\tx{B}^{}N/6$.  
The variation of $c$ gives
\aln{
-s^2+\frac{1}{\tilde{\lambda}_\tx{B}^{}}c-\int\frac{d^dp_\tx{E}^{}}{(2\pi)^d}\frac{1}{p_\tx{E}^2+m_\tx{B}^2+c-i\epsilon}=0,
\label{eom of c in ssb}
}
while the variation of $s$ gives
\aln{(-\Box+m_\tx{B}^2+c)s=0.  
}
As long as we focus on the ground state, we can omit $\Box s$.   
For a given value of $\tilde{\lambda}_\tx{B}^{}$, the parameter space of $m_\tx{B}^{2}$ is divided into two regions as shown in Fig.~\ref{fig:large N}, which correspond to the broken phase (blue) and the unbroken phase (uncolored) respectively.  
Now let us discuss the behavior of the partition function in each region.

\

\noindent {\bf Broken phase}\\ 
When $s\neq 0$, we have $m_\tx{B}^2+c=0$. 
Eq.~(\ref{eom of c in ssb}) then becomes
\aln{
s^2=-\frac{1}{\tilde{\lambda}_\tx{B}^{}}m_\tx{B}^2-\int^\Lambda \frac{d^dp_\tx{E}^{}}{(2\pi)^d}\frac{1}{p_\tx{E}^2}>0,  
\label{s solution}
}
which indicates the parameter space of broken phase. 
This is shown by the blue region in Fig.~\ref{fig:large N}.  
%
%
We represent the critical value of $m_\tx{B}^2$ as $-m_\Lambda^{2}$. 
For $d=4$, it is 
\aln{
m_\Lambda^2=\frac{\tilde{\lambda}_\tx{B}^{}\Lambda^2}{16\pi^2}~.
}
For the parameters obeying Eq.~(\ref{s solution}), the canonical partition function Eq.~(\ref{effective theory in large N}) becomes very simple in the large $N$ analysis. 
In fact, 
\aln{
e^{im_B^2NA}Z\left(m_\tx{B}^2\right)&\sim \exp\left(N\left\{{\rm const}.+im_\tx{B}^2A+\frac{i}{4\tilde{\lambda}_\tx{B}^{}}\int d^dx c^2\right\}\right)
\\
&=\exp\left[N\left\{{\rm const}.+i\frac{V_d^{}}{4\tilde{\lambda}_\tx{B}^{}}\left(m_\tx{B}^2+\frac{2\tilde{\lambda}_\tx{B}^{}A}{V_d^{}}
\right)^2\right\}\right]~.
\label{free energy when SSB}
} 
%
One can see that there is a saddle point at $m_\tx{B}^2=-2\tilde{\lambda}_\tx{B}^{}A/V_d^{}$
if it is smaller than the critical value $-m_\Lambda^2$.   
Otherwise, the exponent in Eq.~(\ref{free energy when SSB}) is a monotonic function for $m_\tx{B}^2\leq -m_\Lambda^2$. 
In particular, the second derivative is given by
\aln{\frac{\partial^2 \log Z(m_\tx{B}^2)}{\partial (m_\tx{B}^2)^2}=i\frac{NV_d^{}}{2\tilde{\lambda}_\tx{B}^{}}~.
\label{second derivative in broken phase}
}

\

\noindent {\bf Unbroken phase}\\
When $s=0$, the system is in the unbroken phase.  
In this case, $c$ is determined by the gap equation 
\aln{
c=\tilde{\lambda}_\tx{B}^{}\int\frac{d^dp_\tx{E}^{}}{(2\pi)^d}\frac{1}{p_\tx{E}^2+m_\tx{B}^2+c-i\varepsilon}.
\label{gap equation}
}
This equation has a solution only for 
\aln{-\frac{1}{\tilde{\lambda}_\tx{B}^{}}m_\tx{B}^2-\int_0^\Lambda\frac{d^dp_\tx{E}^{}}{(2\pi)^d}\frac{1}{p_\tx{E}^2}<0,
}
which is consistent with Eq.~(\ref{s solution}). 
In this case, the canonical partition function becomes 
\aln{
Z\!\left(m_\tx{B}^2\right)\sim \exp\left[iNV_d^{}\left\{-\frac{1}{2}\int\frac{d^dp_\tx{E}^{}}{(2\pi)^d}\log (p_\tx{E}^2+m_\tx{B}^2+c-i\varepsilon)+\frac{1}{4\tilde{\lambda}_\tx{B}^{}}c^2
\right\}
\right],\label{free energy with no SSB}
}
where $c$ depends on $m_\tx{B}^2$ via the gap equation (\ref{gap equation}).  
The exponent in Eq.~(\ref{free energy with no SSB}) is a monotonically decreasing function of $m_\tx{B}^2$ and 
the second derivative with respect to $m_\tx{B}^2$ at the critical point $-m_\Lambda^2$ is 
\aln{\frac{\partial^2 \log Z(m_\tx{B}^2)}{\partial (m_\tx{B}^2)^2}\bigg|_{m_\tx{B}^2=-m_\Lambda^2}^{}&=\frac{iNV_d^{}}{2}\left[\int \frac{d^dp_\tx{E}^{}}{(2\pi)^d}\frac{1}{(p_\tx{E}^2+m_\tx{B}^2+c)^2}\left(1+\frac{dc}{dm_\tx{B}^2}\right)^2+\frac{1}{\tilde{\lambda}_\tx{B}^{}}\left(\frac{dc}{dm_\tx{B}^2}\right)^2
\right]\bigg|_{m_\tx{B}^2=-m_\Lambda^2}^{}
\nn
&=\frac{iNV_d^{}}{2 \tilde{\lambda}_\tx{B}^{}}
\frac{\tilde{\lambda}_\tx{B}^{}\int \frac{d^dp_\tx{E}^{}}{(2\pi)^d}\frac{1}{(p_\tx{E}^2)^2}}{1+\tilde{\lambda}_\tx{B}^{}\int \frac{d^dp_\tx{E}^{}}{(2\pi)^d}\frac{1}{(p_\tx{E}^2)^2}}~.
\label{second derivative in unbroken phase}
}
By comparing Eq.~(\ref{second derivative in broken phase}) and Eq.~(\ref{second derivative in unbroken phase}), one can see that the second derivative is discontinuous at the critical point $m_\tx{B}^2=-m_\Lambda^2$.  
%
Thus, we conclude that $m_\tx{B}^2$ is fixed at $-m_\Lambda^2$ by the mathematical formula~(\ref{general discontinuity}) (as long as $A\neq 0$) at which the renormalized mass $m_\tx{R}^2=m_\tx{B}^2+c$ is zero.
%
%
%
%
%
%

\section{Fixing quartic coupling in $\phi^4$ theory}\label{quartic coupling tuning section}
We briefly comment on the possibility of fixing the quartic coupling. 
As well as the mass term, we can also consider the generalized partition function for the quartic coupling 
\aln{
\int dm_\tx{R}^2\int d\lambda_\tx{R}^{}
\omega(m_\tx{R}^{2},\lambda_\tx{R}^{})e^{im_\tx{R}^{2}A_{m^2}^{}+i\lambda_\tx{R}^{}A_\lambda^{}}Z(m_\tx{R}^2,\lambda_\tx{R}^{}),
}
where we take the renormalized couplings as the integration parameters instead of the bare couplings for simplicity.\footnote{
The Jacobian from the change of variables has been absorbed by the redefinition of $\omega$.
}
Note that $A_\lambda^{}$ is another extensive parameter proportional to the spacetime volume $V$. 
For simplicity, let us restrict the parameter space to $\lambda_\tx{R}^{}\geq 0$ to guarantee the stability of the system. 

At the one-loop level of the $\phi^4$ theory, the vacuum energy~(\ref{one-loop vacuum energy}) vanishes at the critical point of the  mass-squared $m_\tx{R}^{2}=0$ for any $\lambda_\tx{R}^{}\geq 0$, which means that the above partition function becomes 
\aln{
\sim \int_{0}^\infty d\lambda_\tx{R}^{} \omega(\lambda_\tx{R}^{})e^{i\lambda_\tx{R}^{}A_\lambda^{}}. 
}
As long as $A_\lambda^{}\neq 0$, the exponent is a linear function of $\lambda_\tx{R}^{}$, and this integration is strongly dominated by $\lambda_\tx{R}^{}=0$ by the formula Eq.~(\ref{kink delta}) in the large volume limit.  
%
Thus, the free scalar theory seems to be realized in the generalized QFT in the $\phi^4$ theory at least at one-loop level.   
As long as the renormalized mass $m_\tx{R}^2$ is fixed at zero and we focus on the cut-off independent part of the vacuum energy, this conclusion will not be significantly altered by the higher-loop effects because the cut-off independent part should be proportional to $(m_\tx{R}^2)^2$ by dimensional analysis and they vanish at the critical point $m_\tx{R}^{2}=0$. 
%

On the other hand, non-trivial saddle point can appear if we also include the cut-off dependent parts. 
For example, at the three-loop level, we have the following contributions to the vacuum energy:
\aln{\lambda_{\rm R}^{}\frac{A_\lambda^{}}{V_4^{}}+(c_1^{}\lambda_{\rm R}^{}-c_2^{}\lambda_{\rm R}^2) \Lambda^4~,
}
where $c_i^{}>0$ are just constants containing the loop suppression factors.  
%
%
Thus, one can see that there exists a saddle point at 
\aln{
\lambda_\tx{R}^{}\sim \left(\frac{A_\lambda^{}}{V_4^{}}+c_1^{}\right)\bigg/c_2^{}~.
}
%
This is a very interesting possibility but its physical meaning is subtle as in the mass parameter case. 
%
More detailed studies are left for future investigations. 
See also Appendix~\ref{Appendix large N} for the possibility of fixing $\lambda_\tx{R}^{}$ in the large $N$ limit.

%

%

%

\section{Dimensional transmutation in two-real-scalar model}\label{sec:5}
As a next non-trivial example, we study a two-real-scalar model~\cite{Haruna:2019zeu,Hamada:2020wjh,Kannike:2020qtw,Kawai:2021lam,Hamada:2021jls,Hamada:2022soj} at the one-loop level.  
We show that an automatic tuning of the mass-squared parameter is realized so that the dimensional transmutation is successfully achieved.

For simplicity, we again impose the $\mathbb{Z}_2^{}\times  \mathbb{Z}_2^{}$ symmetry, which leads to the following Lagrangian 
\aln{
{\cal L}=-\frac{1}{2}(\partial \phi)^2-\frac{1}{2}(\partial S)^2-\frac{m_\phi^2}{2}\phi^2-\frac{m_S^2}{2}S^2-\frac{\lambda_{\phi S}^{}}{4}\phi^2 S^2-\frac{\lambda_\phi^{}}{4!}\phi^4-\frac{\lambda_S^{}}{4!}S^4.
}
To simplify the discussion further, we focus on the parameter space $\lambda_\phi^{}\ll \lambda_S^{}\sim 1$ and $\lambda_{\phi S}^{}>0$, which means that the $\phi$ direction is almost flat for $m_\phi^{2}\sim 0$. 
$\phi$ and $S$ play the roles of the scalar and gauge fields in the original Coleman-Weinberg mechanism. 
%
%
On the other hand, the $S$ direction is always dominated by  the tree-level potential, that is, its VEV is well determined by $m_{S}^2 S^2+\frac{\lambda_S^{}}{4!}S^4$.

Under these conditions, we will show that there is a critical point at $m_S^2=0$ and $m_\phi^{2}\neq 0$ which corresponds to a quantum first-order phase transition point. 
Although the conventional classical conformal point $m_S^2=m_\phi^2=0$ is not  realized in the current formulation, our result serves as a concrete way of dimensional transmutation without assuming an artificial symmetry such as classical conformality.  
%

Now let us discuss the details. 
The one-loop effective potential in the $\overline{\rm MS}$  scheme is given by
\aln{
V_{\rm eff}^{}(\phi,S)=\frac{m_\phi^2}{2}\phi^2&+\frac{m_S^2}{2}S^2+\frac{\lambda_{\phi S}^{}}{4}\phi^2 S^2+\frac{\lambda_\phi^{}}{4!}\phi^4+\frac{\lambda_S^{}}{4!}S^4
\nn
&+\frac{(M_+^{2}(\phi,S))^2}{64\pi^2}\log \left(\frac{M_+^{2}(\phi,S)}{\mu^2 e^{3/2}}\right)+\frac{(M_-^{2}(\phi,S))^2}{64\pi^2}\log \left(\frac{M_-^{2}(\phi,S)}{\mu^2 e^{3/2}}\right),
}
where
\aln{M_\pm^2(\phi,S)=\frac{1}{2}\left(M_\phi^{2}+M_S^2\pm \sqrt{(M_\phi^{2}-M_S^2)^2+4\lambda_{\phi S}^2\phi^2S^2}\right)
}
with
\aln{M_\phi^2=m_\phi^2+\frac{\lambda_{\phi S}^{}}{2}S^2+\frac{\lambda_\phi^{}}{2}\phi^2,\quad M_S^2=m_S^2+\frac{\lambda_{\phi S}^{}}{2}\phi^2+\frac{\lambda_S^{}}{2}S^2. 
\label{BG dependen masses}
}
Here, all the coupling constants are the renormalized ones. 
%
%
%
As usual, we can take the renormalization scale at the point $\mu=M$ where $\lambda_\phi^{}$ vanishes.  
In the following, we first examine how $m_\phi^2$ is fixed for a given value of $m_S^2$ and then discuss the fixing of $m_S^2$. 
%

\begin{figure}
\begin{center}
\includegraphics[width=10cm]{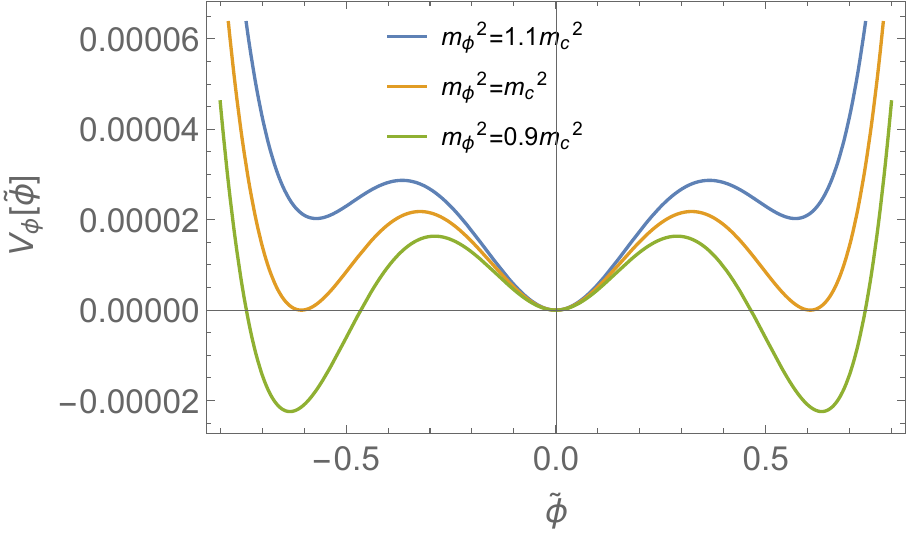}
\end{center}
\caption{One loop effective potential of $\tilde{\phi}$. 
Different colors correspond to the different values of $m_\phi^2$. 
}
\label{fig:Vphi}
\end{figure}

First, let us consider the parameter space $m_S^{2}\geq 0$. 
%
In this case, $\langle S\rangle=0$ as long as the tree-level potential dominates in the $S$ direction. 
%
The effective potential for $\phi$ is then
\aln{
V_{\rm eff}^{}(0,\phi)&=
\frac{(m_\phi^2)^2}{64\pi^2}\log \left(\frac{m_\phi^2}{M^2 e^{3/2}}\right)+\frac{m_\phi^2}{2}\phi^2+\frac{
(m_S^2+\frac{\lambda_{\phi S}^{}}{2}\phi^2)^2}{64\pi^2}\log \left(\frac{m_S^2+
\frac{\lambda_{\phi S}^{}}{2}\phi^2}{M^2 e^{3/2}}\right)~.
\label{two-scalar effective potential}
} 
In Fig.~\ref{fig:Vphi}, we show the plots of $V_{\rm eff}^{}(0,\phi)$ where the different colors correspond to the different values of $m_\phi^2$. 
In particular, one can see that there exists a quantum first-order phase transition at 
\aln{m_\phi^2=\frac{\lambda_{\phi S}^{}e^{1/2}}{64\pi^2}M^2=:m_c^2~,
\label{critical mc}
}
meaning that $\phi$ achieves a nonzero VEV, $\langle \phi\rangle=v_\phi^{}$, for $m_\phi^{2}<m_c^2$.  
See Appendix~\ref{app:two scalar} for the derivation of Eq.~(\ref{critical mc}).  
Correspondingly, the VEV and the vacuum energy are given by
\aln{
\langle \phi \rangle
	&=	\begin{cases} 0 & {\rm for}~m_\phi^2\geq m_c^2 
\\
v_\phi^{} & {\rm for}~m_\phi^2< m_c^2
\end{cases}~,
\\
V_{\rm eff}^{}(0,\langle \phi\rangle)-\frac{(m_\phi^2)^2}{64\pi^2}\log \left(\frac{m_\phi^2}{M^2 e^{3/2}}\right)
	&= \begin{cases} \frac{
(m_S^2)^2}{64\pi^2}\log \left(\frac{m_S^2}{M^2 e^{3/2}}\right) & {\rm for}~m_\phi^2\geq m_c^2
\\
\frac{m_\phi^2}{2}v_\phi^2
+\frac{
(m_S^2+\frac{\lambda_{\phi S}^{}}{2}v_\phi^2)^2}{64\pi^2}\log \left(\frac{m_S^2+
\frac{\lambda_{\phi S}^{}}{2}v_\phi^2}{M^2 e^{3/2}}\right)
& {\rm for}~m_\phi^2< m_c^2
\end{cases}~.
}
As in the simple $\phi^4$ theory, the additional contribution in the second line gives the discontinuity of the second derivative of the vacuum energy at $m_\phi^{2}=m_c^2$, and this point is dominant in the generalized partition function. 

Second, we turn to the case $m_S^{2}<0$. 
In this case, $S$ already has a VEV at $\langle S\rangle^2=-6m_S^2/\lambda_S^{}$ at tree level, and there exists the tree-level vacuum energy $-3(m_S^2)^2/(2\lambda_S^{})$ as in the simple $\phi^4$~theory discussed in the previous sections.       
As for the $\phi$ potential, nonzero VEV of $S$ just changes the effective mass in Eq.~(\ref{BG dependen masses}) as 
\aln{M_S^2=m_S^2+\frac{\lambda_{S}^{}}{2}\langle S\rangle^2+\frac{\lambda_{\phi S}^{}}{2}\phi^2=2|m_S^2|+\frac{\lambda_{\phi S}^{}}{2}\phi^2.
}
Note that $M_\phi^2$ is just a constant because of $\lambda_\phi^{}=0$. 
Thus, we can repeat the same discussion as above 
and conclude that $m_\phi^2$ is still fixed at the critical point $m_\phi^2=m_c^2$.   

Finally, at such a critical point of $m_\phi^2$, the vacuum energy as a function of $m_S^2$ is given by
\aln{
V_{\rm min}^{}\bigg|_{m_\phi^2=m_c^2}\simeq 
\begin{cases} \frac{
(m_S^2)^2}{64\pi^2}\log \left(\frac{m_S^2}{M^2 e^{3/2}}\right)   & \text{for $m_S^{}\geq 0$}
\\
-\frac{3(m_S^2)^2}{2\lambda_S^{}} +\frac{
(m_S^2+\frac{\lambda_{\phi S}^{}}{2}v_\phi^2)^2}{64\pi^2}\log \left(\frac{m_S^2+
\frac{\lambda_{\phi S}^{}}{2}v_\phi^2}{M^2 e^{3/2}}\right) & \text{for $m_S^{}<0$}
\end{cases}~.
\label{Vmin two scalar}
}    
As in the simple $\phi^4$ theory
, $m_S^2$ is fixed at zero because the second derivative of this equation is discontinuous at $m_S^{2}=0$.  
%
The conventional Coleman-Weinberg point $m_S^2=m_\phi^2=0$ is not realized in the current formation because such a point $m_S^2=m_\phi^2=0$ corresponds to the degeneracy of false vacua as discussed in Ref.~\cite{Kawai:2021lam}. 
We leave the discussion on the criticality of general extrema for future investigation.  
%

\section{Summary}\label{sec:6}
In this paper, we have investigated the Higgs hierarchy problem in the generalized QFT.        
We first studied the free scalar theory and found that there are two critical points in the large volume limit: One is $m_\tx{B}^2=0$ and the other is $m_\tx{B}^2={\cal O}(\Lambda^2)$. 
While the former does not depend on the regularization methods, the latter does, implying that $m_\tx{B}^2=0$ is a physically reasonable critical point. 
This can be a theoretical origin/explanation of classical conformality which is implicitly assumed in many literatures.

We then studied the $\phi^4$ theory at the one-loop level, as well as the large $N$ model. 
In this case, we found that a critical point exists at the point where the renormalized mass $m^2=m_{\tx{B}}^2+\delta m_{\rm UV}^2$ vanishes due to the discontinuity of the vacuum energy's derivative.  
%
This further backs up the fine-tuning of the (Higgs) mass-squared being automatically accomplished in the generalized QFT. 
%
%

We have also discussed the possibility that the quartic coupling is automatically fixed, with a positive result.

As a next non-trivial example, we examined the $\mathbb{Z}_2^{}$ invariant two-real-scalar model at the one-loop level by focussing on a simple parameter space where only one real scalar $\phi$ can develop a nonzero VEV $\langle \phi\rangle\neq 0$.   
Under these conditions, we found that 
there exists a critical point where $m_\phi^{2}=m_c^2>0$ and $m_S^2=0$, which corresponds to a quantum first-order phase transition point of the theory.  
%
%
This result gives a complete realization of dimensional transmutation without assuming classical conformality which has been implicitly assumed in many literatures.  
%
%

\section*{Acknowledgements} 
The work of KK is supported by KIAS Individual Grants, Grant No. 090901
The work of KO is partly supported by the JSPS Kakenhi Grant Nos.~19H01899 and 21H01107.  
The work of KY is supported in part by the Grant-in-Aid for Early-Career Scientists, No.~19K14714. 
H.K. thanks Prof. Shin-Nan Yang and his family for their kind support through the Chin-Yu chair professorship. H.K. is partially supported by JSPS (Grants-in-Aid for Scientific Research Grants No.20K03970), by the Ministry of Science and Technology, R.O.C. (MOST 111-2811-M-002-016), and by National Taiwan University

\appendix 
\section*{Appendix}

\section{Mathematical formulas}\label{sec:Preliminaries}
In the following, we often use mathematical formulas of  generalized functions. 
We summarize them here. 
The first one is well known as the saddle point approximation: 
\aln{
\lim_{V\rightarrow \infty}e^{-Vg(x)}\simeq e^{-Vg(x_0^{})}\sqrt{\frac{2\pi}{Vg''(x_0^{})}}\times \delta(x). 
\label{saddle point delta}
}
where $g(x)$ is a smooth function and $x_0^{}$ is a saddle point, $g'(x_0^{})=0$. 
The proof of this equation is textbook level and we will not repeat it here. 
%
The second formula is 
\aln{
\lim_{V\rightarrow \infty}e^{iVg(x)}\theta(x)\simeq \frac{i}{V}\left|\frac{dg}{dx}\right|^{-1}e^{iVg(0)}\times \delta(x),
\label{kink delta}
}
where $g(x)$ is now a real, smooth and monotonic function  for $x\geq 0$ and satisfies $g'(0)\neq 0$. 
%
%
The proof is as follows. 
By multiplying a test function $f(x)$ with a finite support and integrating from $0$ to $\infty$, we have
\aln{
\int_0^\infty dx e^{iV g(x)}f(x)&=\int_{g(0)}^{g(\infty)} dg\left|\frac{dg}{dx}\right|^{-1} e^{iVg}f(x=x(g))
\\
&=\left[\frac{e^{iVg}}{iV}\left|\frac{dg}{dx}\right|^{-1}f(x(g))\right]_{g(0)}^{g(\infty)}-\frac{1}{iV}\int_{g(0)}^{g(\infty)} dg\frac{d}{dg}\left(\left|\frac{dg}{dx}\right|^{-1}f(x(g))\right) e^{iVg}
\nn
&=i\frac{e^{iVg(0)}}{V}\left|\frac{dg}{dx}\right|^{-1}f(x)\bigg|_{x=0}^{}-\frac{1}{(iV)^2}\left[\frac{d}{dg}\left(\left|\frac{dg}{dx}\right|^{-1}f(x(g))\right) e^{iVg}\right]_{g(0)}^{g(\infty)}+\cdots
\nn
&=i\frac{e^{iVg(0)}}{V}\left|\frac{dg}{dx}\right|^{-1}f(x)\bigg|_{x=0}^{}+{\cal O}(V^{-2}), 
}
which confirms Eq.~(\ref{kink delta}).   
Note that the contribution from $x=\infty$ is zero because we have assumed that $f(x)$ has a finite support. 
More generally, when $g(x)$ is a monotonic and smooth function for both sides $x>0$ and $x<0$, but its derivative $g'(x)$ is discontinuous at $x=0$, we have  
\aln{
\lim_{V\rightarrow \infty}e^{iVg(x)}=\lim_{V\rightarrow \infty}\frac{i}{V}\left[\left|\frac{dg}{dx}\right|^{-1}_{0+}-\left|\frac{dg}{dx}\right|^{-1}_{0-}\right]e^{iVg(0)}\times \delta(x).
\label{two sided equation}
}
Note that when $g'(x)$ is also continuous at $x=0$, the right-hand side vanishes and higher order terms dominate. 
In general, when the derivatives of $g(x)$ are continuous and nonzero up to $(n-1)$-th order but the $n$-th derivative of $g(x)$ is discontinuous, we have 
\aln{\lim_{V\rightarrow \infty}e^{iVg(x)}=c\lim_{V\rightarrow \infty}\frac{i^n}{V^n}
e^{iVg(0)}\times \delta(x),
\label{general discontinuity}
}
where $c$ is a coefficient determined by $g'(0),g''(0),\cdots,d^{(n-1)}g(0)/dx^{n-1},dg^{(n)}(0+)/dx^{n},dg^{(n)}(0-)/dx^{n}$.   
\section{Fixing quartic coupling in large $N$ model}\label{Appendix large N}
In this Appendix, we study a possibility of fixing the quartic coupling by taking the $N^{-1}$ corrections.
In the following, the mass term $m^2$ is fixed at the critical point $m^2=m_\tx{B}^2+c_{\rm cl}^{}=0$ and we focus on $d=4$.  

Around the large $N$ solution $(c,s)=(c_{\rm cl}^{},s_{\rm cl}^{})$ studied in Section~\ref{large N}, the effective action in Eq.~(\ref{effective theory in large N}) becomes 
\aln{
S=\frac{N\Lambda^{4}V_4^{}}{4\tilde{\lambda}_\tx{B}^{}}\left(\frac{\tilde{\lambda}_\tx{B}^{}}{16\pi^2}\Lambda^2\right)^2+N\int d^4x\left(-\frac{1}{2}(\partial_\mu^{}\delta s)^2-\frac{1}{2}\delta c\delta s^2+\frac{1}{4}\left(\frac{1}{\tilde{\lambda}_\tx{B}^{}}+\int \frac{d^4p_\tx{E}^{}}{(2\pi)^4}\frac{1}{(p_\tx{E}^2)^2}\right)\delta c^2+\cdots
\right),
\label{large N action}
}
where $\delta s=s-s_{\rm cl}^{},~\delta c=c-c_{\rm cl}^{}$ and $\cdots$ represents the higher order terms. 
Note that the first term is the leading vacuum energy contribution and it is proportional to $\tilde{\lambda}_\tx{B}$. 
Thus, this can be absorbed into the definition of $A_\lambda^{}$.   
In the following, we represent 
\aln{
\frac{1}{\tilde{\lambda}_\tx{B}^{}}+\int \frac{d^4p_\tx{E}^{}}{(2\pi)^4}\frac{1}{(p_\tx{E}^2)^2}=\frac{1}{\tilde{\lambda}_\tx{B}^{}}+\frac{1}{8\pi^2}\log\left(\frac{\Lambda}{\mu}\right):=\frac{1}{N\lambda(\mu)^{}}, 
\label{lambdas}
}
where $\mu$ is some (IR) scale. 
We can check that $\lambda(\mu)$ corresponds to the usual renormalized quartic coupling because Eq.~(\ref{lambdas}) is nothing but the solution of RGE in the large $N$ limit:
\aln{\frac{d\lambda}{d\log \mu}\approx \frac{N}{8\pi^2}\lambda^2. 
}
%
%
In Eq.~(\ref{large N action}), $\delta c$ part can be rewritten as
\aln{\frac{1}{4N\lambda^{}(\mu)}\left(\delta c-N\lambda^{}(\mu)\delta s^2\right)^2-\frac{N\lambda^{}(\mu)}{4}\delta s^4. 
}
%
By the field redefinitions $N^{1/2}\delta s\rightarrow \delta s$ and $N^{1/2}(\delta c-N\lambda^{}(\mu)\delta s^2)\rightarrow \delta c$, the action becomes
\aln{
S=
\int d^4x\left(-\frac{1}{2}(\partial_\mu^{}\delta s)^2-\frac{\lambda^{}(\mu)}{4}\delta s^4+\frac{1}{4N\lambda(\mu)}\delta c^2+{\cal O}(N^{-1})
\right),
} 
%
%
Now we can perform the one-loop integration of $\delta c$ as 
\aln{-\frac{V_4^{}}{2}\int \frac{d^4q_\tx{E}^{}}{(2\pi)^4}\log(N\lambda(\mu)/2)^{-1}
=-\frac{F}{2}\log\left(\frac{2}{N\lambda(\mu)}\right) =-\frac{F^{}}{2}\log\left(\frac{1}{\lambda_\tx{B}^{}}+\frac{N}{8\pi^2}\log\left(\frac{\Lambda}{\mu}\right)\right)+{\rm const}, 
}
where $F^{}=V_4^{}\Lambda^4/(32\pi^2)$. 
Note that we have singular points at
\aln{
\lambda(\mu)=0,\quad  \pm\infty.\label{singular points}
} 
Now the $\lambda_\tx{B}^{}$ integration in the micro-canonical partition function is given by
\aln{
\Omega(A)&=\int_{-\infty}^\infty \frac{d\lambda_\tx{B}^{}}{2\pi}\exp\left(i\lambda_\tx{B}^{}NA_\lambda^{}-i\frac{F^{}}{2}\log\left(\frac{1}{\lambda_\tx{B}^{}}+\frac{N}{8\pi^2}\log \left(\frac{\Lambda}{\mu}\right)\right)\right)
\\
&=\int_{-\infty}^\infty \frac{d\lambda_\tx{B}^{}}{2\pi}e^{iNA_\lambda^{}G(\lambda^{}(\mu))},
}
where 
\aln{
G(\lambda(\mu)):=\frac{\lambda(\mu)}{1-\frac{N\lambda(\mu)}{8\pi^2}\log\left(\frac{\Lambda}{\mu}\right)}+\frac{F^{}}{2NA_\lambda^{}}\log\lambda(\mu)~
.
}
One can check that there exist two saddle points:
\aln{
N\lambda^{}(\mu)=N\lambda_\pm^{}=\frac{1}{B}\left(1-\frac{A_\lambda^{}}{FB}\pm\sqrt{\left(1-\frac{A_\lambda^{}}{FB}\right)^2-1}\right) ,
} 
where
\aln{B=\frac{1}{8\pi^2}\log\left(\frac{\Lambda}{\mu}\right). 
}
We can see that $N\lambda_{\pm}^{}$ is real and positive if we take the following large $N$ limit. 
\aln{
N\rightarrow \infty~\quad N\lambda_\tx{R}^{}(\mu)={\rm fixed},\quad \frac{A_\lambda^{}}{F}={\rm fixed},\quad \bigg|1-\frac{A_\lambda^{}}{F B}\bigg|\geq 1~.
}
More detailed analysis is necessary to verify the validity of this  limit because we now have additional parameter $A_\lambda^{}$ that is absent in the usual large $N$ analysis.   

\section{Details of two-real-scalar model}\label{app:two scalar}
By putting $\tilde{\phi}^2=(m_S^2+\lambda_{\phi S}^{}\phi^2/2)/M^2e^{3/2}$, we can rewrite Eq.~(\ref{two-scalar effective potential}) as
\aln{
V_{\rm eff}^{}(0,\phi)
&=
\Lambda(m_\phi^2,m_S^2)+\tilde{V}_\phi^{}(\tilde{\phi}),
}
where 
\aln{\Lambda(m_\phi^2,m_S^2)
	&:=	\frac{(m_\phi^2)^2}{64\pi^2}\log \left(\frac{m_\phi^2}{\mu^2 e^{3/2}}\right)-\frac{m_S^2m_\phi^2}{\lambda_{\phi S}},\\
\tilde{V}_\phi^{}(\tilde{\phi})
	&:=	(M^2e^{3/2})^2\left(\frac{\tilde{m}_\phi^2}{2}\tilde{\phi}^2+\frac{
			\tilde{\phi}^4}{64\pi^2}\log\tilde{\phi}^2\right),
}
in which
\aln{
\tilde{m}_\phi^2
	&:=	\frac{2m_\phi^2}{\lambda_{\phi S}^{}M^2e^{3/2}}.
}
The effective potential has minima at $\phi=0$
and 
\aln{\tilde{\phi}=\tilde{v}_\phi^{}~,
}
where $\tilde{v}_\phi^2$ is a solution of 
\aln{\tilde{m}_\phi^2+\frac{\tilde{\phi}^2}{16\pi^2}\log (\tilde{\phi}^2e^{1/2})=0. 
}
Correspondingly, $v_\phi^{}$ will denote the VEV of the original field $\phi$ below. 
%

Note that $\langle \phi\rangle=0$ is always the true vacuum for $m_S^{2}>\frac{\lambda_{\phi S}^{}}{2}v_\phi^2$ because $\tilde{\phi}^2>\tilde{v}_\phi^2$ for any values of $\phi^2>0$. 
%
On the other hand, the minimum of $\tilde{V}_\phi^{}(\tilde{\phi})$ depends on $m_\phi^2$ for $\frac{\lambda_{\phi S}^{}}{2}v_\phi^2\geq m_S^2\geq 0$.  
In particular, it has a critical point at
\aln{
0=\tilde{V}_{\phi}^{}(0)=\tilde{V}_{\phi}^{}(\tilde{v}_\phi^{})\quad \rightarrow \quad \tilde{m}_\phi^2=\frac{1}{32\pi^2e^{}}=:\tilde{m}_c^2~.
}
See Fig.~\ref{fig:Vphi} for the explicit plots of $\tilde{V}_\phi(\tilde{\phi})$. 
%

As a consistency check of $\langle S\rangle=0$ for $m_S^2\geq 0$, let us also check the positivity of the effective mass of $S$ at $S=0$:
\aln{m_{{\rm eff},S}^2&=\frac{\partial V_{\rm eff}^{}}{\partial S^2}\bigg|_{S=0,\phi=\langle \phi\rangle}^{}
\\
&=m_S^2+\frac{\lambda_{\phi S}^{}}{2}\langle\phi\rangle^2+\frac{\lambda_S^{}}{32\pi^2}\left(m_S^{2}+\frac{\lambda_{\phi S}^{}}{2}\langle\phi\rangle^2\right)\log\left(\frac{m_S^2+\frac{\lambda_{\phi S}^{}}{2}\langle\phi\rangle^2}{M^2e^{}}\right)
+\frac{\lambda_{\phi S}^{}m_\phi^2}{32\pi^2}\log\left(\frac{m_\phi^2}{M^2e^{}}\right).\label{effective S mass}
}
The last term is negligible at around the critical point $m_\phi^2=m_c^2$ because  
\aln{
\frac{\lambda_{\phi S}^{}m_\phi^2}{32\pi^2}\log\left(\frac{m_\phi^2}{M^2e^{}}\right)\sim -\frac{\lambda_{\phi S}^2}{(32\pi)^2}\langle \phi\rangle^2\ll \frac{\lambda_{\phi S}^2}{2}\langle \phi\rangle^2.
}
On the other hand, 
the second term in Eq.~(\ref{effective S mass}) seems to become negative at around $m_S^2\sim 0$ for $\langle \phi\rangle=0$, but it is just an illusion of looking at the first term of leading log corrections. 
%
By summing up all the leading log terms (which corresponds to the RG improvement), we have
\aln{1+\frac{\lambda_S^{}}{32\pi^2}\log\left(\frac{m_S^2}{M^2e^{}}\right)+\left(\frac{\lambda_S^{}}{32\pi^2}\log\left(\frac{m_S^2}{M^2e^{}}\right)\right)^2+\cdots =\frac{1}{1-\frac{\lambda_S^{}}{32\pi^2}\log\left(\frac{m_S^2}{M^2e^{}}\right)},
}
which is always positive for $m_S^2\geq 0$.\footnote{This is nothing but the well-known fact~\cite{Coleman:1973jx} that the Coleman-Weinberg mechanism does not work for a single (real) scalar.} 
Thus, $\langle S\rangle=0$ is justified at this one-loop level calculations and $m_\phi^2$ is fixed at the critical point $m_c^2$ by the discontinuity of the vacuum energy.    

\bibliography{Bibliography}
\bibliographystyle{JHEP}

\end{document}